\documentclass[twocolumn,amssymb,floatfix,aps,prl]{revtex4}
\usepackage{url}
\usepackage{graphicx}
\begin{document}                                                
\begin{titlepage}
\begin{flushright}BNL-HET-04/12
\end{flushright}
\begin{flushright}FSU-HEP-2004/0804
\end{flushright} 
\begin{flushright}UB-HET-04-02
\end{flushright}
\begin{flushright}hep-ph/0408077
\end{flushright}
\vspace{1truecm}
\begin{center}
{\large\bf
Higgs boson production with one bottom quark jet at hadron colliders
}
\\
\vspace{0.5in}
{\bf S.~Dawson}\\
{\it Department of Physics, Brookhaven National Laboratory,\\
Upton, NY 11973-5000, USA}
\\ 
\vspace{.25in}
{\bf C.B.~Jackson, L.~Reina}\\
{\it Department of Physics, Florida State University,\\ 
Tallahassee, FL 32306-4350, USA}
\\
\vspace{.25in}
{\bf D.~Wackeroth}\\
{\it  Department of Physics, State University of New York at Buffalo, \\
Buffalo, NY 14260-1500, USA} 
\vspace{2in}  

{\bf Abstract}
\end{center}
  We present total rates and kinematic distributions for the
  associated production of a single bottom quark and a Higgs boson at the
  Tevatron and the LHC. We include next-to-leading order QCD
  corrections and compare the results obtained in the four and five flavor 
  number schemes for parton distribution functions.
\end{titlepage}
\vspace*{5cm}
\newpage
\setcounter{page}{0}
\title{
Higgs boson production with one bottom quark jet at hadron colliders
}
\author{S.~Dawson$^{a}$, C.B.~Jackson$^{b}$, L.~Reina$^{b}$,  and
 D.~Wackeroth$^{c}$}
\bigskip
\affiliation{
a) Department of Physics, Brookhaven National Laboratory,
Upton, NY 11973-5000, USA\\
b) Physics Department, Florida State University, 
Tallahassee, FL 32306-4350, USA\\
c)  Department of Physics, State University of New York,
Buffalo, NY 14260-1500, USA}
\begin{abstract} 
  We present total rates and kinematic distributions for the
  associated production of a single bottom quark and a Higgs boson at the
  Tevatron and the LHC. We include next-to-leading order QCD
  corrections and compare the results obtained in the four and five flavor 
  number schemes for parton distribution functions.
\end{abstract}
\maketitle

One of the most pressing problems of particle physics is to uncover
the origin of electroweak symmetry breaking.  The Standard Model (SM)
predicts the existence of one scalar particle, the Higgs boson, as a
consequence of the generation of gauge boson masses via the
Higgs-Kibble mechanism.  Extensions of the SM often introduce more
Higgs fields, whose properties may drastically differ from the SM
Higgs boson.  Finding experimental evidence for the existence of one
or more Higgs bosons and measuring their couplings to gauge bosons,
leptons, and quarks is a major goal of particle physics.

In this letter, we focus on Higgs boson production in association with
bottom ($b$) quarks.  The coupling of the Higgs boson to a $b$ quark
is suppressed in the SM by the small factor $m_b/v$, where
$v\!=\!(\sqrt{2}G_F)^{-1/2}\!=\!246$~GeV, implying that associated
production of a SM Higgs boson with $b$ quarks is very small at both
the Tevatron and the LHC.  In a two Higgs doublet model or a
supersymmetric model, however, this coupling is proportional to the
ratio of neutral Higgs boson vacuum expectation values, $\tan\beta$,
and can be significantly enhanced for large values of $\tan\beta$.

\begin{figure}[hbt]
\begin{center}
\includegraphics[scale=0.8]{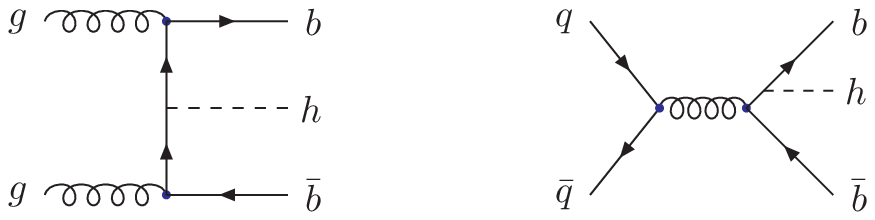}
\vspace*{-0.2cm}
\caption[]{Sample Feynman diagrams for $gg\rightarrow b\bar{b}h$ and 
  $q\bar{q}\rightarrow b\bar{b}h$ production at tree level.}
\label{fg:ggbbh_feyn}
\end{center}
\end{figure}

\begin{figure}[hbt]
\begin{center}
\includegraphics[scale=0.8]{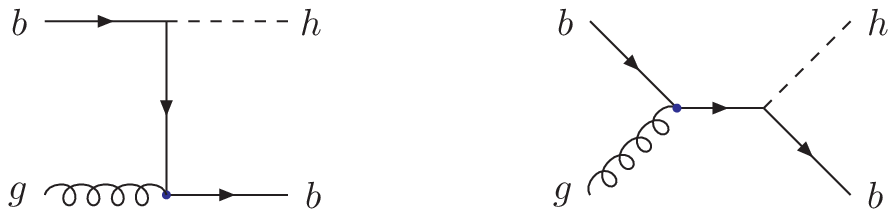}
\vspace*{-0.2cm}
\caption[]{Feynman diagrams for $gb\rightarrow bh$ production at tree level.}
\label{fg:bgbh_feyn}
\end{center}
\end{figure}

The associated production of a Higgs boson with $b$ quarks proceeds at
tree level via $q\bar{q},gg\to b\bar{b}h$, as illustrated in
Fig.~\ref{fg:ggbbh_feyn}. Fully inclusive or semi-inclusive Higgs
production cross sections are then obtained by imposing no
identification cuts on the final state $b$ quarks or by requiring that
at least one $b$ quark is observed at high transverse momentum, $p_T$,
respectively.  This approach with no $b$ quarks in the initial state
is dubbed the fixed or \emph{four flavor number scheme} (4FNS).  Large
logarithms of the form $\Lambda\!=\!\log(\mu_h^2/m_b^2)$ (for
$\mu_h\!\approx M_h$) arise at all orders in $\alpha_s$ from the
integration over the $p_T$ of the final state $b$ quarks that
originate from the collinear splitting of an initial state gluon.  In
order to stabilize the perturbative expansion of the corresponding
cross section, these logarithms can be resummed into a $b$ quark
perturbatively defined parton distribution function (PDF), which
integrates over the low $p_T$ region of the corresponding $b$ quark up
to scales of the order of the factorization scale. This approach is
identified as the variable or \emph{five flavor number scheme}
(5FNS) \cite{Barnett:1987jw,Olness:1987ep,Dicus:1989cx}.  The 5FNS is
based on the approximation that the outgoing $b$ quarks are at small
$p_T$. It can be used only if the corresponding $b$ quark is treated
inclusively and no $p_T$ cut is required to identify the $b$ quark in
the final state. In the 5FNS the leading processes for fully inclusive
and semi-inclusive Higgs production with $b$ quarks are $b\bar{b}\to
h$ and $bg\to bh$, respectively (the case of $bg\to bh$ is illustrated
in Fig.~\ref{fg:bgbh_feyn}, at tree level). In this approach the
processes $q\bar{q},gg\to b\bar{b}h$ are subleading contributions of
higher order in the $\alpha_s^n \Lambda^m$
expansion \cite{Dicus:1998hs,Balazs:1998sb}.

Assessing the validity and compatibility of the 4FNS and 5FNS
approaches in the context of Higgs production with $b$ quarks has
recently been the subject of much theoretical interest. In this
letter, we focus exclusively on the production of a Higgs boson with
one identified $b$ quark jet, and analyze in detail the results
obtained within the 4FNS and 5FNS.  Requiring one final state $b$
quark measures unambiguously the Yukawa coupling of the $b$ quark and
significantly enhances the rate with respect to the case when both $b$
quarks are identified. Higgs boson production with one $b$ quark jet
(and $h\to b\bar{b}$) has been extensively studied by the CDF and D0
collaborations \cite{Affolder:2000rg,bbh_d0} and is going to play a
major role in the experimental searches for Higgs bosons beyond the SM
at the Tevatron and at the LHC. Thus, a more dedicated effort aimed at
refining the theoretical predictions for both total and differential
cross sections is mandatory.

\begin{figure}[hbt]
\begin{center}
\includegraphics[scale=0.5]{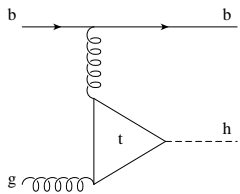}
\vspace*{-0.2cm}
\caption[]{Feynman diagram for the closed top quark loop
contribution to $gb\rightarrow bh$.}
\label{fg:bgbhtriangle}
\end{center}
\end{figure}

A first study of the next-to-leading order (NLO) QCD total cross
sections for $q\bar{q},gg\to b\bar{b}h$
\cite{Dittmaier:2003ej,Dawson:2003kb} and for $bg\to bh$
\cite{Campbell:2002zm} processes has been presented in
Ref.~\cite{Campbell:2004pu}.  In this letter we concentrate on the
comparison of total and differential cross sections at NLO QCD in the
4FNS and 5FNS schemes.  This is the first comparison of differential
cross sections in the two PDF schemes and is important to assess the
residual theoretical uncertainties in future experimental analyses.
In particular, we discuss the effects of including the closed top
quark loop diagram of Fig.~\ref{fg:bgbhtriangle}, a contribution that
had been previously neglected, in the NLO calculation of $bg\to bh$ in
the 5FNS.

The NLO QCD corrections to $pp,p\bar p \to b(\bar b)h$ production in
the 4FNS consist of calculating the ${\cal O}(\alpha_s)$ virtual and
real QCD corrections to the $q\bar{q},gg\to b\bar{b}h$ tree level
processes \cite{Dittmaier:2003ej,Dawson:2003kb}, imposing
identification cuts on the transverse momentum and pseudorapidity of
either the $b$ or $\bar{b}$ final state quark (antiquark). Results
from the two existing calculations
\cite{Dittmaier:2003ej,Dawson:2003kb} have been compared and found in
good agreement (see Ref.~\cite{Campbell:2004pu}).  Except for the
identification cuts, the calculation is identical to that for
$t\bar{t}h$ production
\cite{Beenakker:2001rj,Beenakker:2002nc,Reina:2001sf,Reina:2001bc,
  Dawson:2002tg,Dawson:2003zu} with the global replacement of the top
quark by the bottom quark mass ($m_t\to m_b$) and Yukawa coupling
($g_{t\bar{t}h}\to g_{b\bar{b}h}$), where $m_b$ is always non zero.
\begin{figure}[hbtp!]
\begin{center}
\includegraphics[scale=0.38]{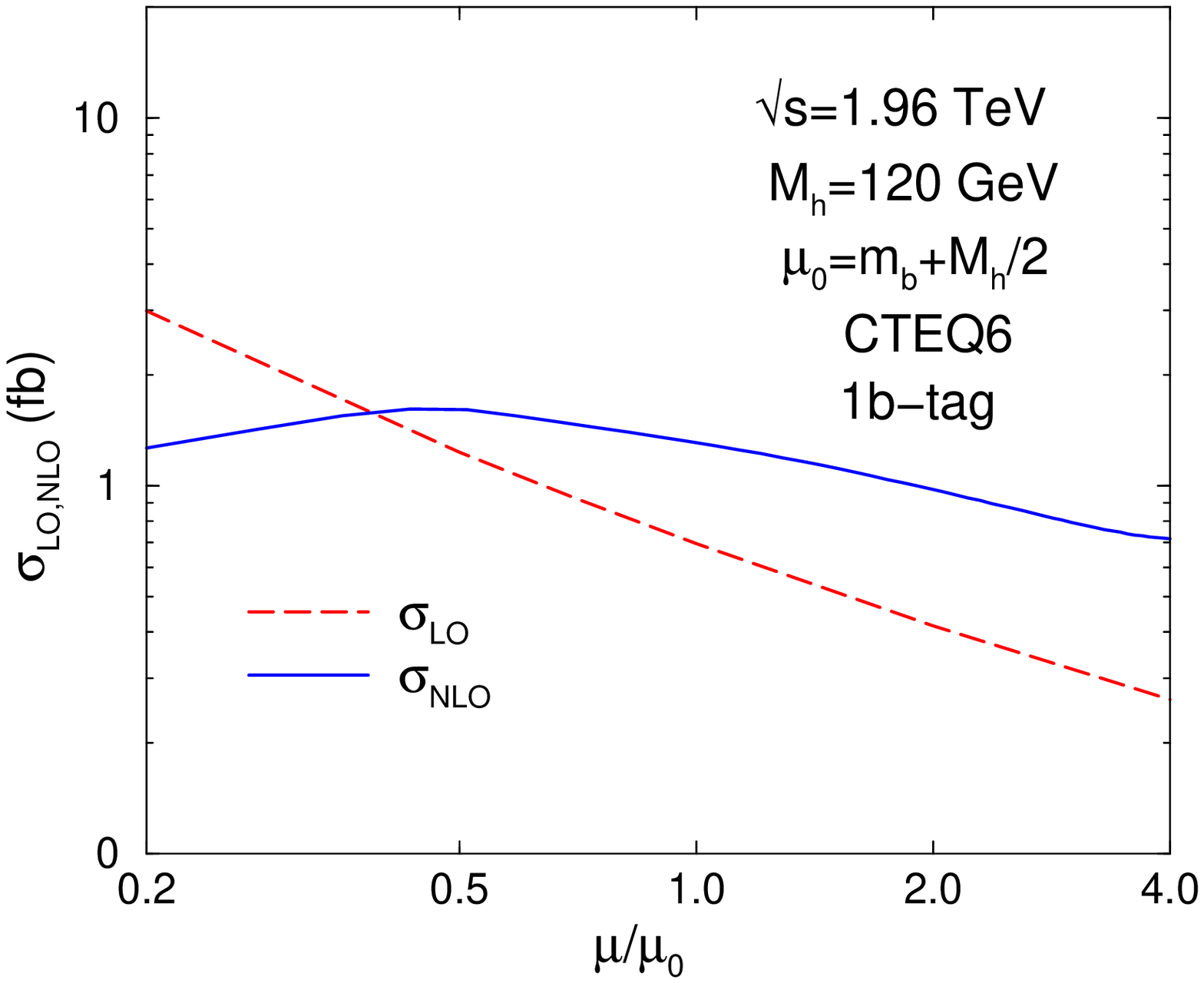} 
\includegraphics[scale=0.38]{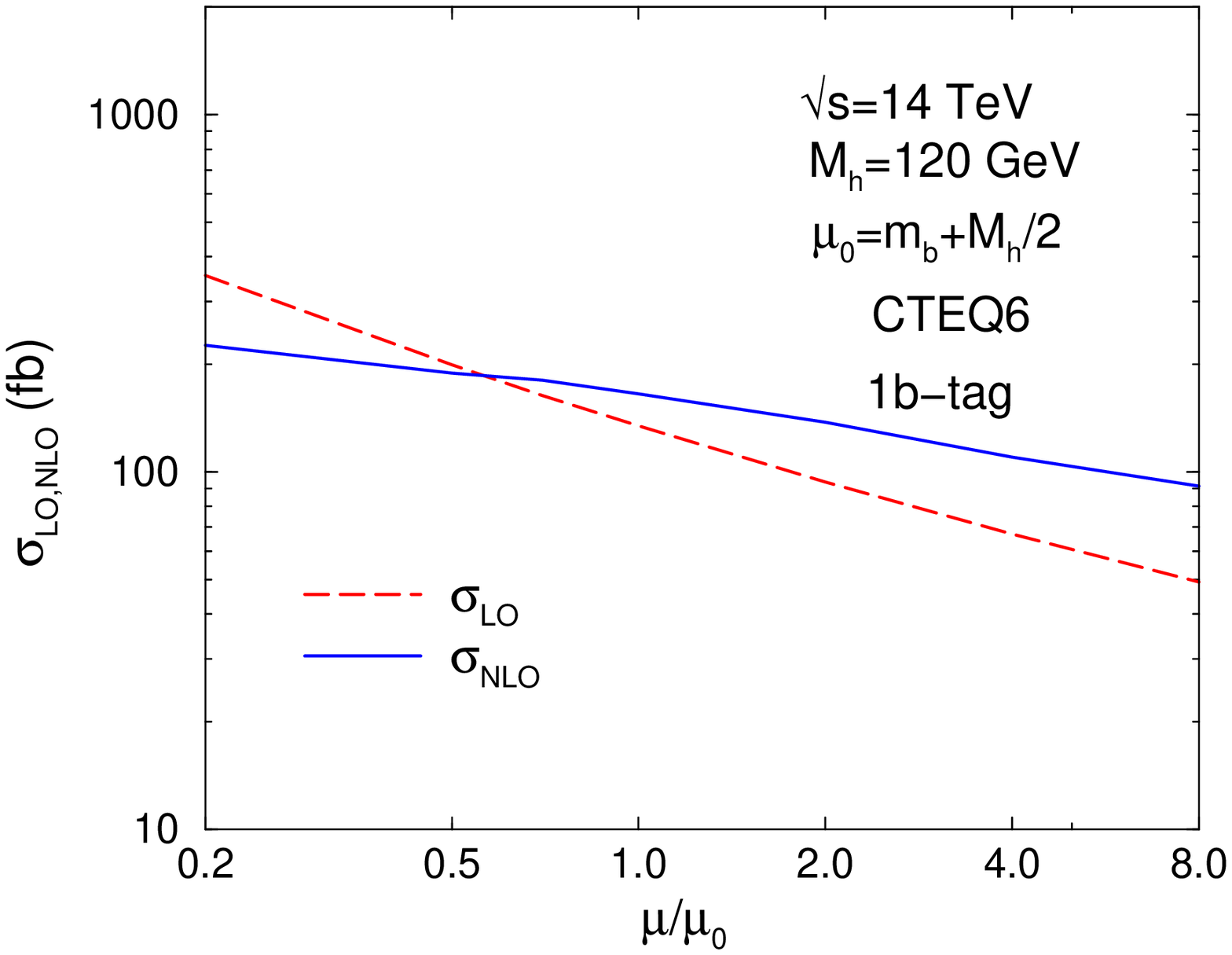}
\vspace*{-0.4cm}
\caption[]{
  Total LO and NLO cross sections for $pp,p\bar p\to b(\bar b) h$
  production in the 4FNS as a function of $\mu\!=\!\mu_r\!=\!\mu_f$
  for $M_h\!=\!120$~GeV, at both the Tevatron and the LHC.}
\label{fg:mudep}
\end{center}
\end{figure}
\begin{figure}[hbtp!]
\begin{center}
\begin{tabular}{c}
\includegraphics[bb=40 30 520 430,scale=0.4]{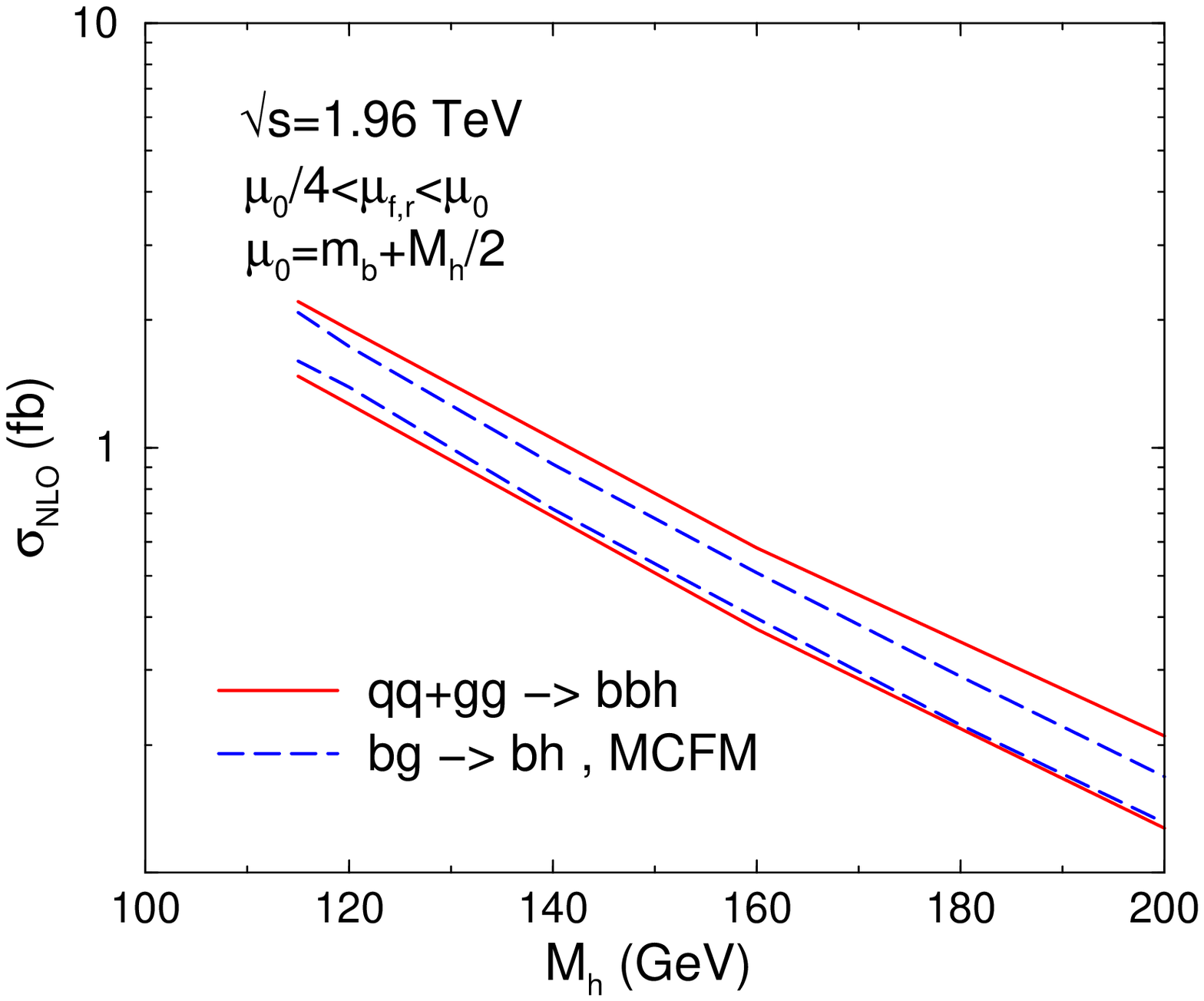}
\includegraphics[bb=600 -440 720 630,scale=0.17]{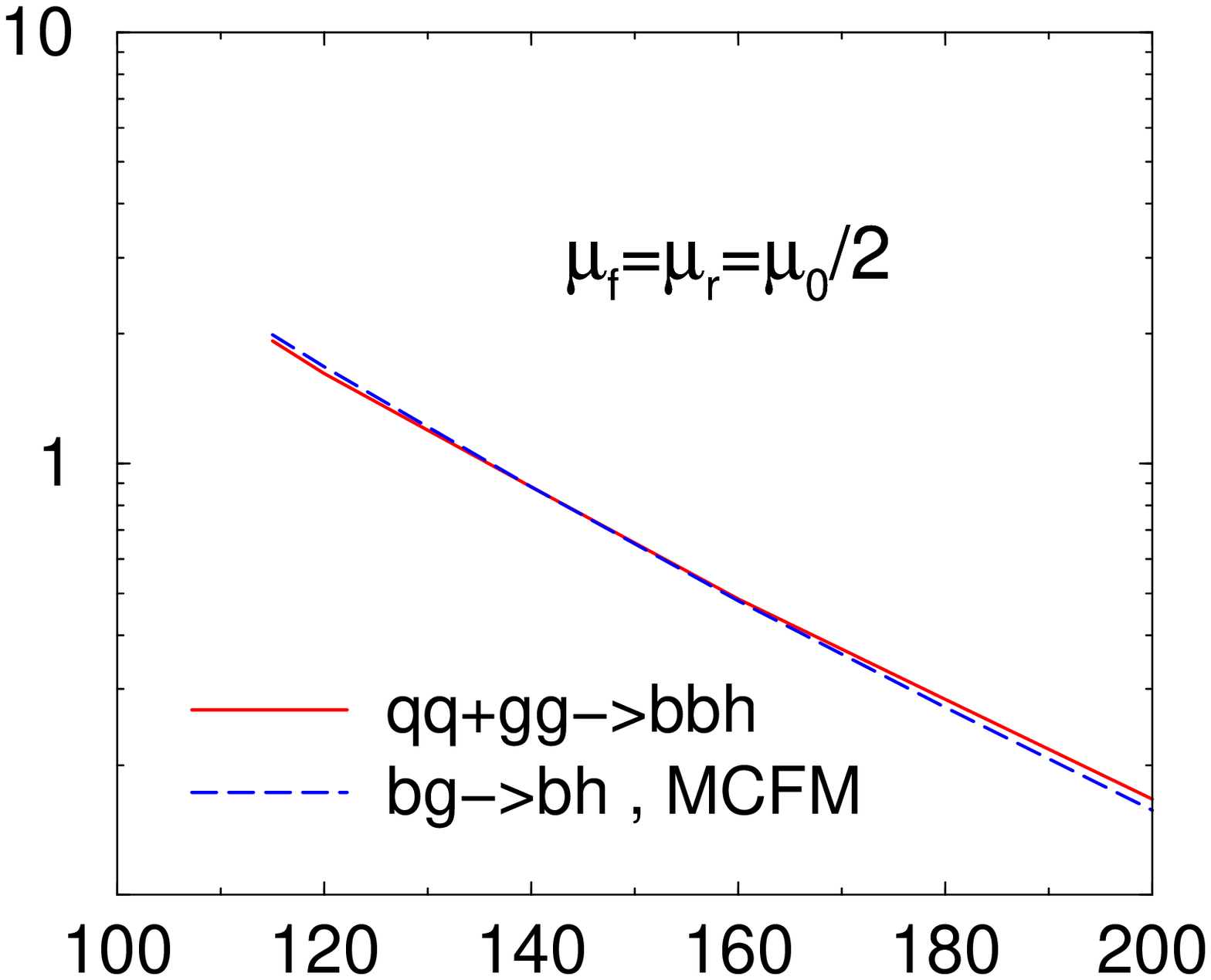}\\
\includegraphics[bb=40 30 520 430,scale=0.4]{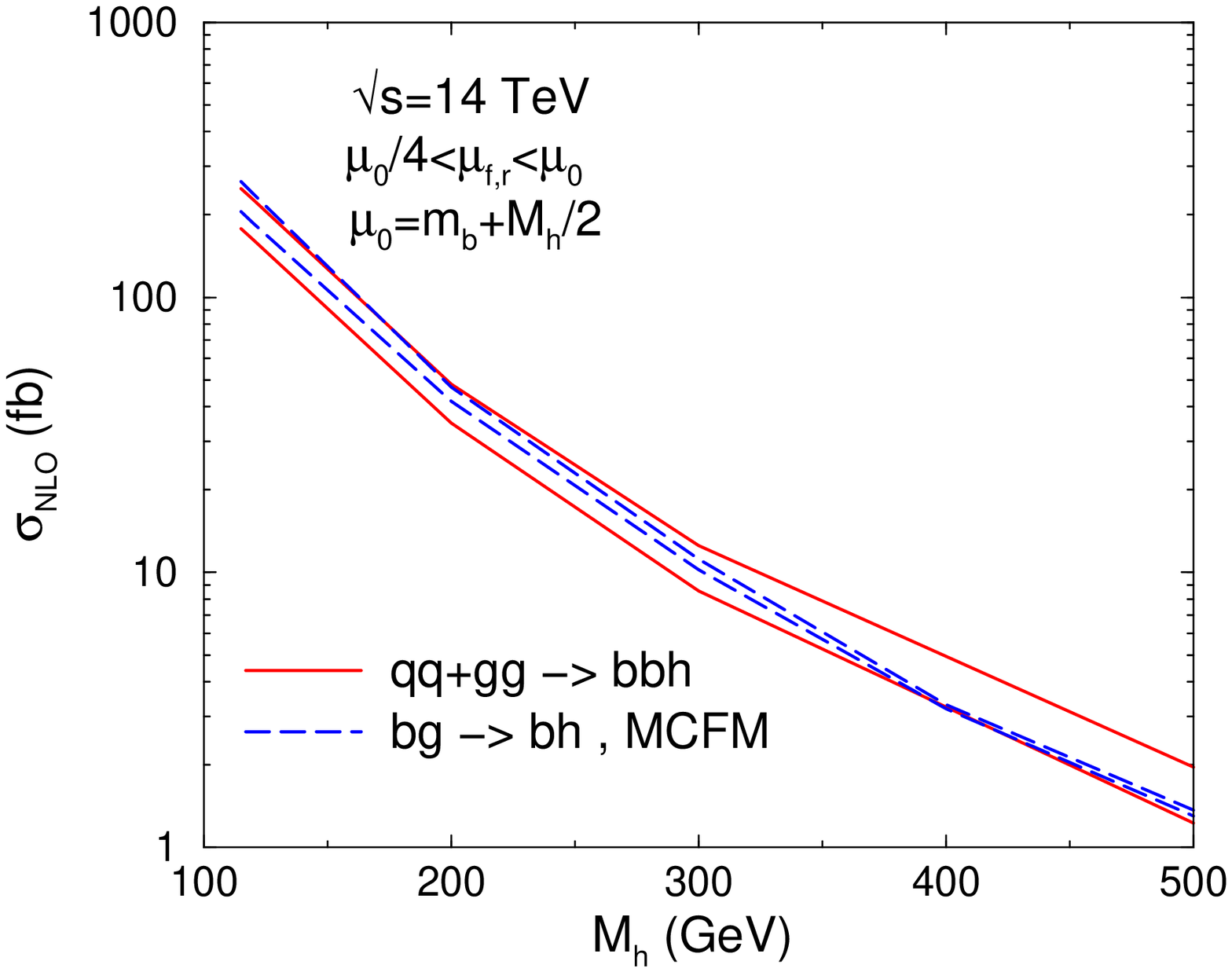}
\includegraphics[bb=600 -440 720 630,scale=0.17]{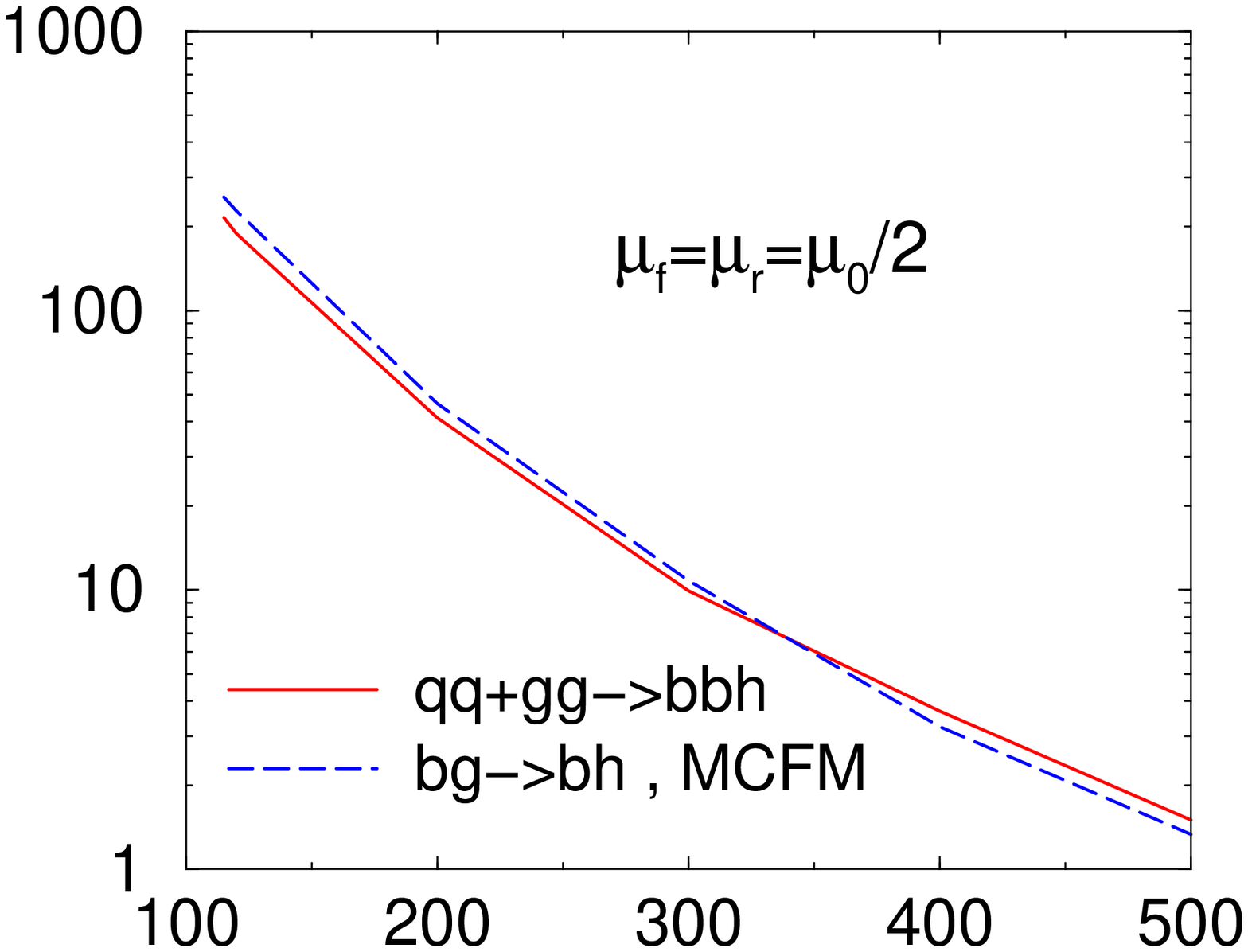}\\
\end{tabular}
\vspace*{-0.4cm}
\caption[]{
  Total NLO cross section for $pp,p\bar p\to b(\bar b) h$ production
  at the Tevatron and the LHC as a function of $M_h$. We have assumed
  $\mu_r\!=\!\mu_f\!=\!\mu_0/2$ for the central curves (see inlays)
  and varied $\mu_r$ and $\mu_f$ independently to obtain the
  uncertainty bands, as explained in the text.  The solid curves
  correspond to the 4FNS, the dashed curves to the 5FNS.}
\label{fg:mhdep}
\end{center}
\end{figure}

The NLO QCD corrections to $pp,p\bar p \to bh+\bar{b}h$ production in
the 5FNS have been presented in Ref.~\cite{Campbell:2002zm} and are
encoded in the Monte Carlo program MCFM \cite{mcfm}. In
Ref.~\cite{Campbell:2002zm}, the calculation of the cross sections for
$bg\to bh$ is performed in the $m_b\!=\!0$ approximation (except for
the $b$ quark Yukawa coupling), and for this reason the only virtual
diagram containing a top quark loop (see Fig.~\ref{fg:bgbhtriangle})
is neglected.  Indeed, the contribution of this diagram to the virtual
cross section is proportional to $g_{t\bar{t}h}g_{b\bar{b}h}m_b/m_t$
and therefore vanishes when $m_b\!=\!0$. At the same time, in the SM
($g_{b\bar{b}h}\!=\!m_b/v$) the contribution of this diagram is
overall of order $g_{b\bar{b}h}^2$ as all other diagrams retained in
the $m_b\!=\!0$ approximation. So, it can play a relevant numerical
role in the comparison between the 5FNS and the 4FNS, where diagrams
with closed top quark loops are included.  To investigate this issue
we have computed this contribution to the $bg\to bh$ process and
implemented it into MCFM.  All numerical results in the 5FNS presented
here are obtained with this modified version of MCFM.

Our LO numerical results are obtained using CTEQ6L1 PDFs
\cite{Pumplin:2002vw,Stump:2003yu} and the 1-loop evolution of
$\alpha_s$, while for NLO results we use CTEQ6M PDFs and the 2-loop
evolution of $\alpha_s$, with $\alpha_s(M_Z)\!=\!0.118$.  We
use the $\overline{MS}$ running $b$ quark mass in the $b$ quark Yukawa
coupling, evaluated at 1- and 2-loops respectively for LO and NLO
results (with pole mass $m_b\!=\!4.62$~GeV). Our renormalization
scheme decouples the top quark from the running of $m_b(\mu)$ and
$\alpha_s(\mu)$ and is explained in detail in
Ref.~\cite{Dawson:2003kb}. We work in the SM but the results can be
straightforwardly generalized to the case of the scalar Higgs bosons
of a supersymmetric extension of the SM, for instance, by replacing
the SM top and bottom quark Yukawa couplings accordingly, as discussed
in Ref.~\cite{Dawson:2003kb}.

In order to simulate the experimental cuts, we require one of the
final state $b$ quarks to have $p_T\!>\!20~$GeV and pseudo-rapidity
$\mid\!\eta\!\mid<\!2.0$ for the Tevatron and $\mid\!\eta\!\mid<\!2.5$
for the LHC. In the NLO real gluon emission, the final state gluon and
$b$ quarks are considered as separate particles only if $\Delta
R\!>\!0.4$ ($\Delta R\!=\!\sqrt{(\Delta\eta)^2+(\Delta\phi)^2}$).

In Fig.~\ref{fg:mudep} we show, for $M_h\!=\!120$~GeV, the dependence
of the LO and NLO total cross sections, calculated in the 4FNS, on the
arbitrary renormalization/factorization scale $\mu$ (with
$\mu_r\!=\!\mu_f\!=\!\mu$).  The NLO result has considerably less
sensitivity to the scale choice, and the region around
$\mu\!\approx\!\mu_0/2$ ($\mu_0\!=\!m_b+M_h/2$) shows the least
sensitivity to the variation of $\mu$. For this reason we use
$\mu_0/2$ as our reference scale in the following plots, whenever
$\mu_r\!=\!\mu_f$.  Analogous results for the 5FNS total cross
sections have been presented in Ref.~\cite{Campbell:2002zm}.

Fig.~\ref{fg:mhdep} shows the dependence of the NLO total cross
sections on $M_h$, in both the 4FNS and 5FNS. The bands illustrate the
theoretical uncertainty due to the independent variation of $\mu_r$
and $\mu_f$ about the central value $\mu_r\!=\!\mu_f\!=\!\mu_0/2$ (see
inlays), between $(0.2,0.25)\mu_0$ (4FNS, 5FNS) and $\mu_0$. It is
extremely interesting to note that the 5FNS band is completely within
the 4FNS band, and the corresponding central values are almost
identical at the Tevatron and very close at the LHC.  Including the
closed top quark loop diagrams lowers the 5FNS cross section (by
$\approx 15\%$ at the Tevatron and $\approx 10\%$ at the LHC, when
$M_h\!=\!120$~GeV and $\mu_r\!=\!\mu_f\!=\!0.5\mu_0$) and makes the
theoretical prediction in the 4FNS and 5FNS fully compatible (see for
comparison Fig.~6 in Ref.~\cite{Campbell:2004pu}). Note that the bands
only give an indication of the theoretical uncertainty of each
approach due to the residual scale dependence.  Other sources of
theoretical uncertainties, like PDF uncertainties, have not been
considered.
\begin{figure}[hbtp!]
\begin{center}
  \includegraphics[bb=150 500 430 715,scale=0.7]{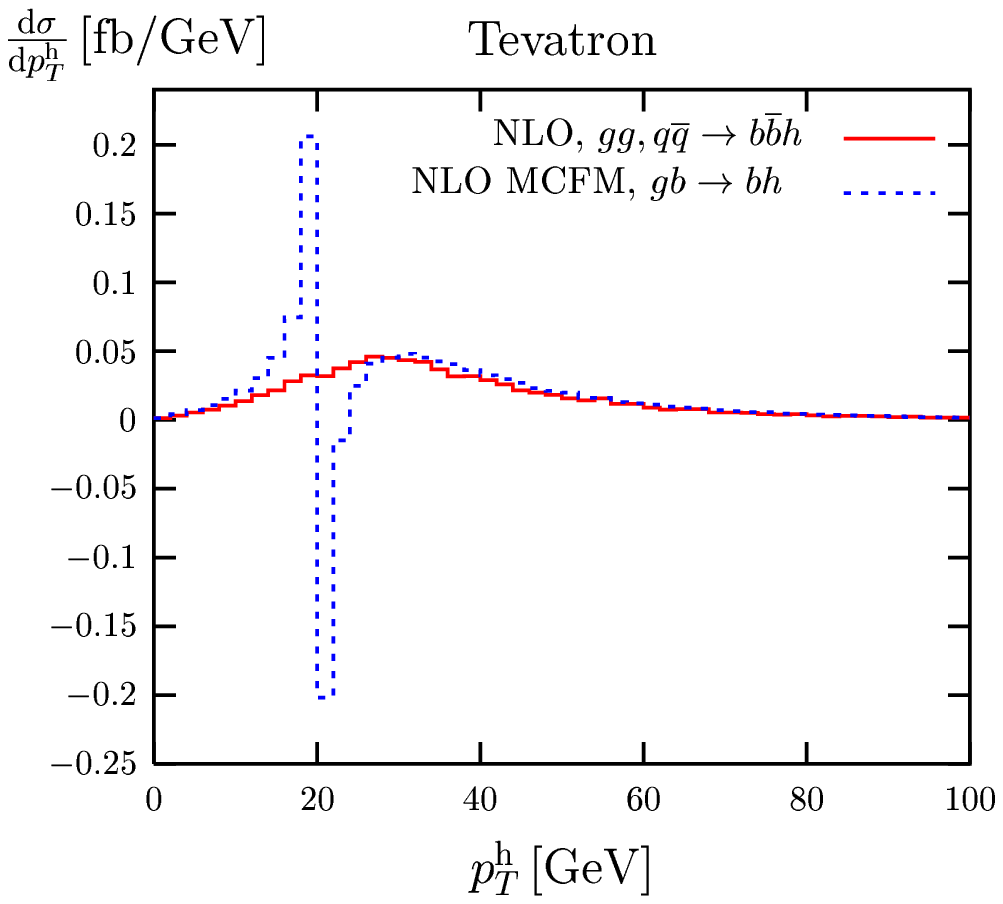}
  \includegraphics[bb=90 460 230 560,scale=0.4]{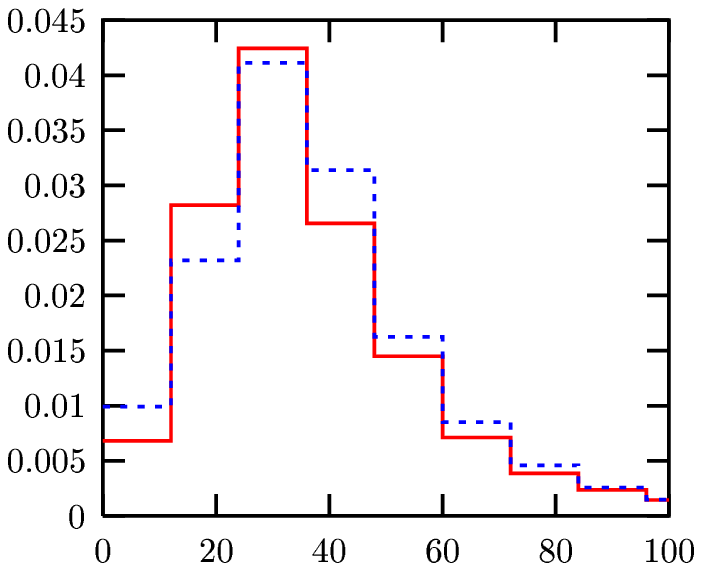}
  \includegraphics[bb=150 500 430 715,scale=0.7]{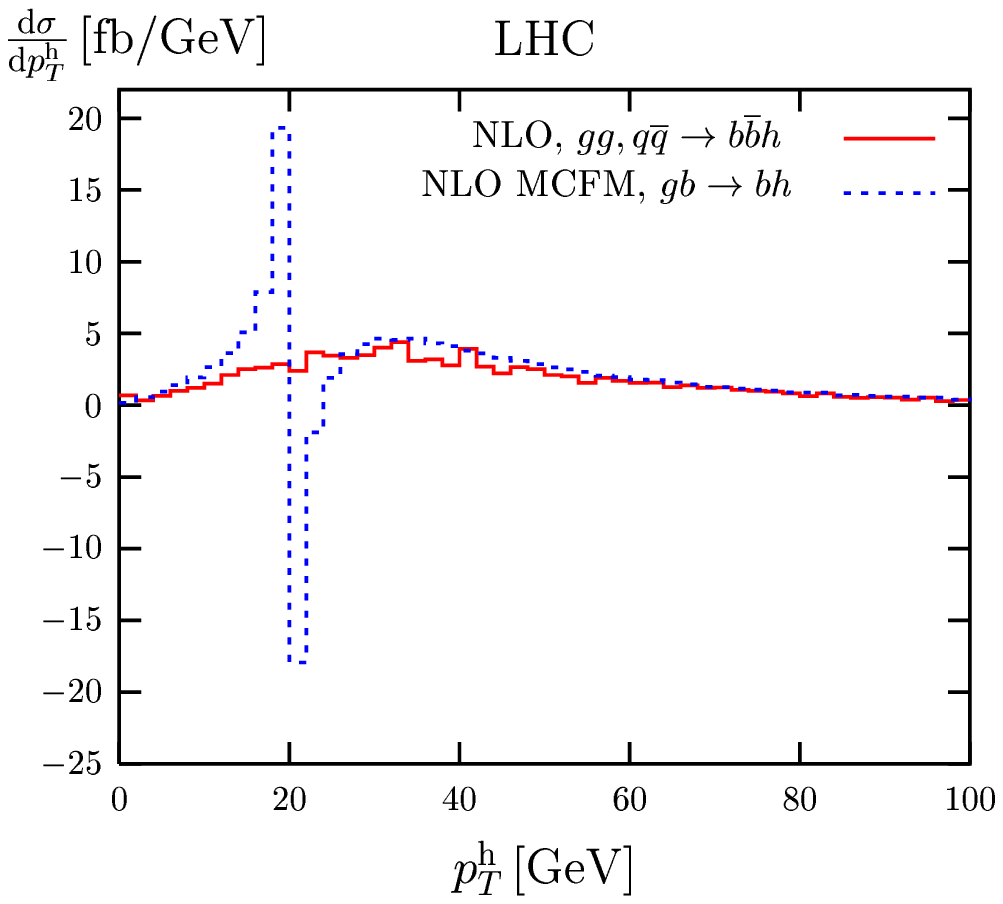}
  \includegraphics[bb=90 460 230 560,scale=0.4]{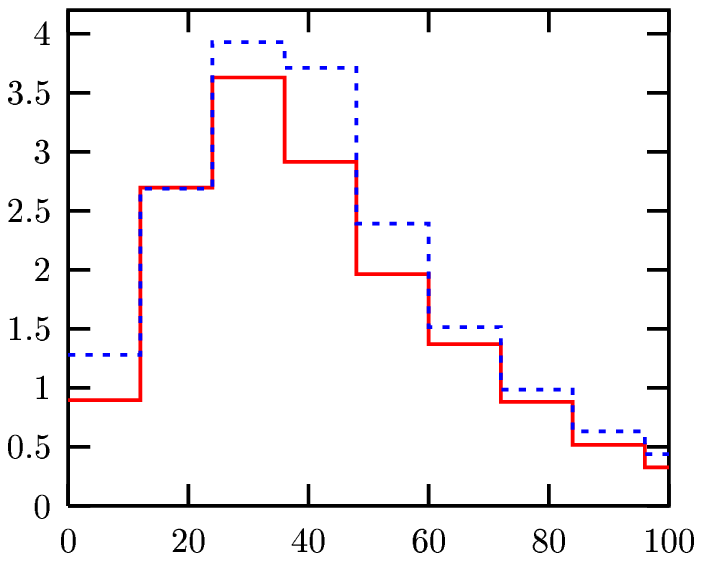}
\vspace*{-0.4cm}
\caption[]{
  $d \sigma/dp_T^h$ at the Tevatron and the LHC for $M_h=120$~GeV and
  $\mu_r=\mu_f=\mu_0/2$.  We show the NLO results in the 4FNS (solid)
  and 5FNS (dashed), using two different bin sizes, 2~GeV and 12~GeV
  (see inlay).}
\label{fg:dsdpt}
\end{center}
\end{figure}

\begin{figure}[hbtp!]
\begin{center}
\begin{tabular}{rl}
\includegraphics[bb=140 450 430 710,scale=0.42]{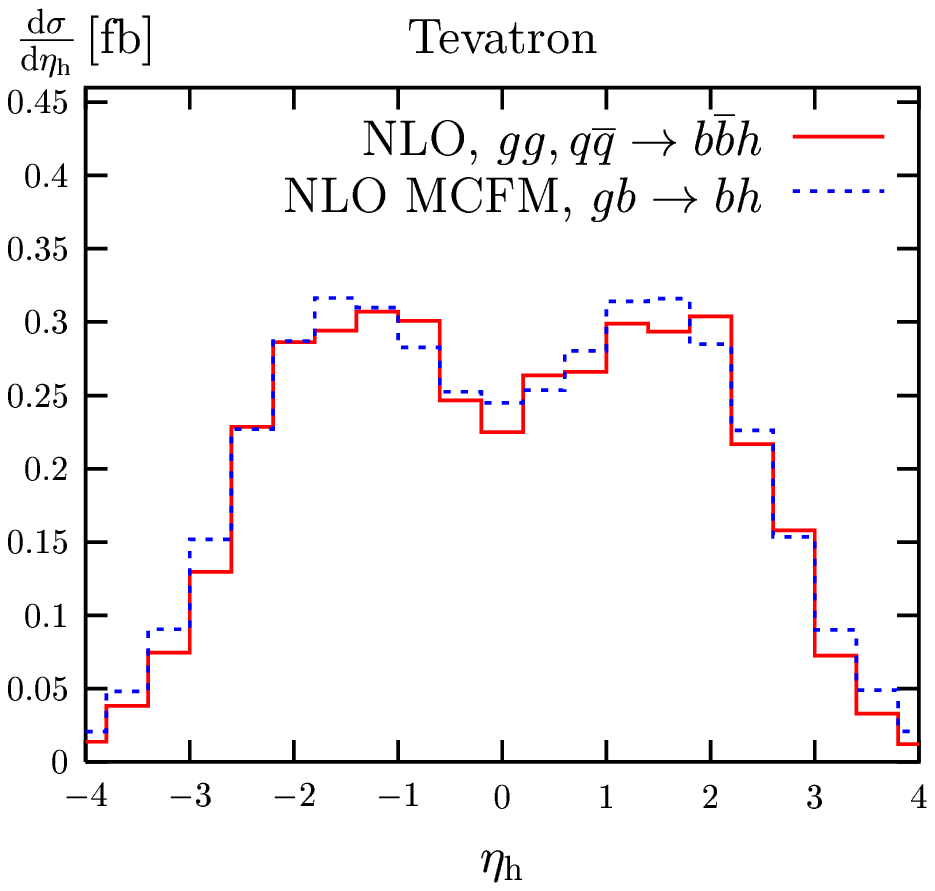} & 
\includegraphics[bb=140 450 430 710,scale=0.42]{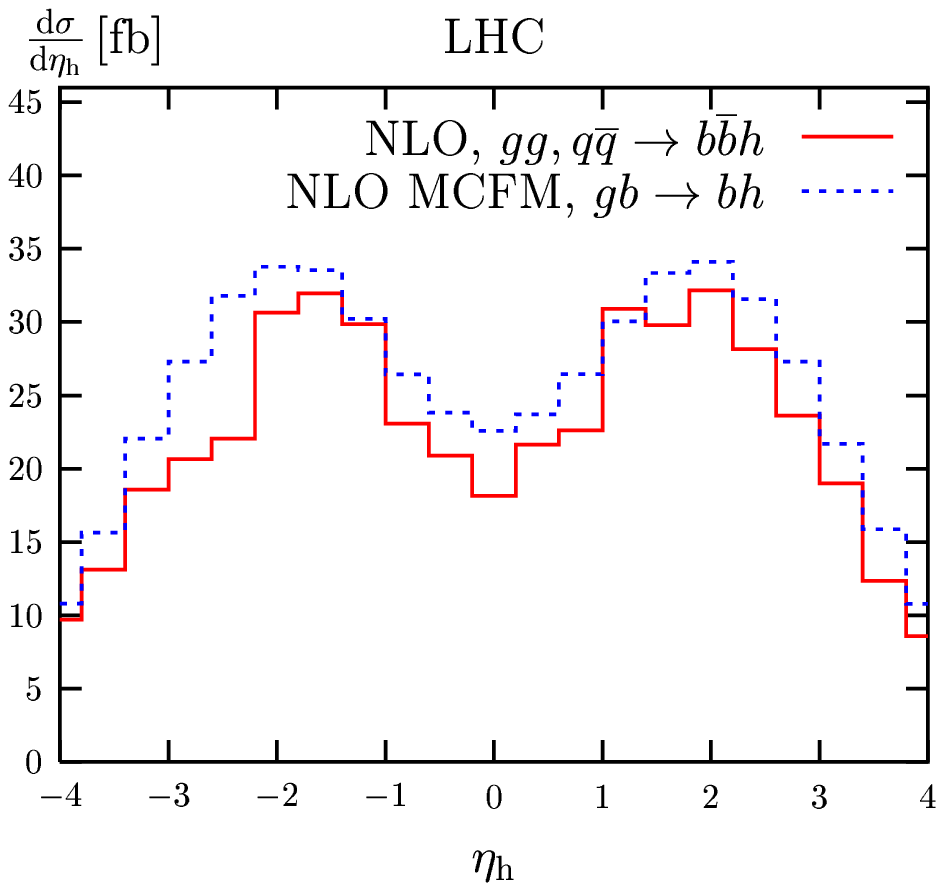}
\end{tabular}
\vspace*{-0.4cm}
\caption[]{
  $d\sigma/d\eta_h$ at the Tevatron and the LHC for $M_h\!=\!120$~GeV
  and $\mu_r\!=\!\mu_f\!=\! \mu_0/2$.  We show the NLO results in the
  4FNS (solid) and 5FNS (dashed).}
\label{fg:dsdeta}
\end{center}
\end{figure}

\begin{figure}[hbtp!]
\begin{center}
\begin{tabular}{cc}
\includegraphics[bb=140 450 430 725,scale=0.42]{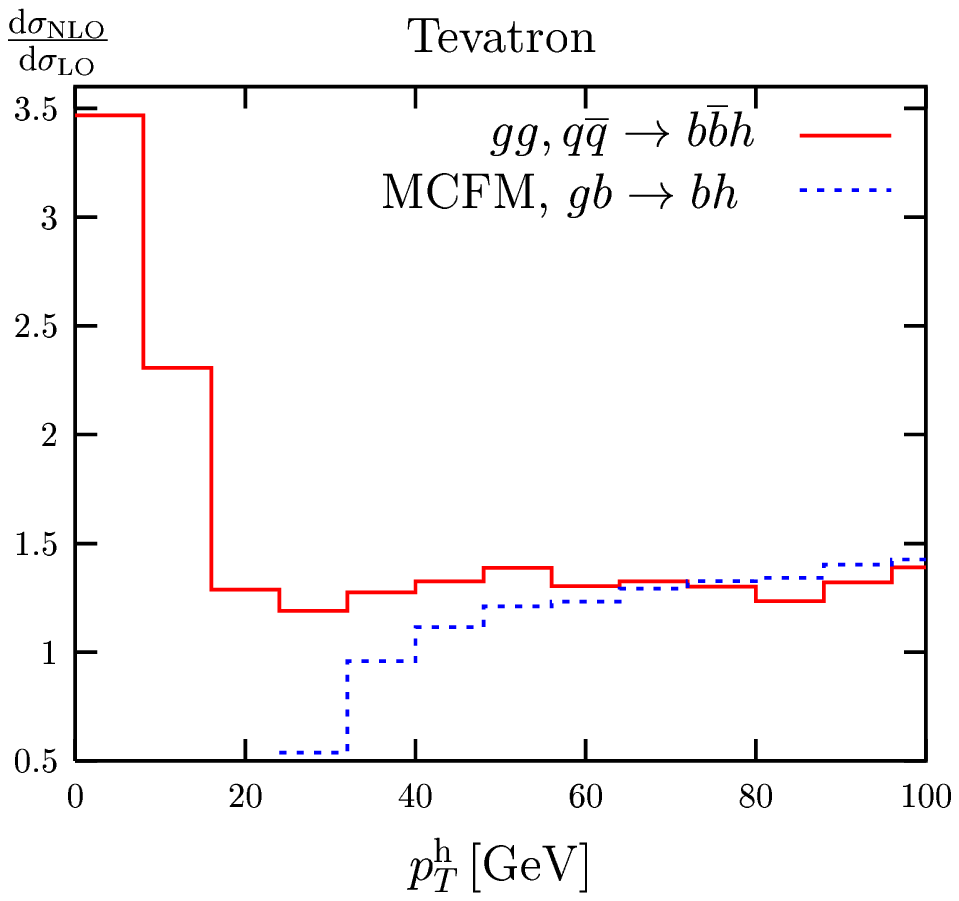}&
\includegraphics[bb=140 455 430 725,scale=0.42]{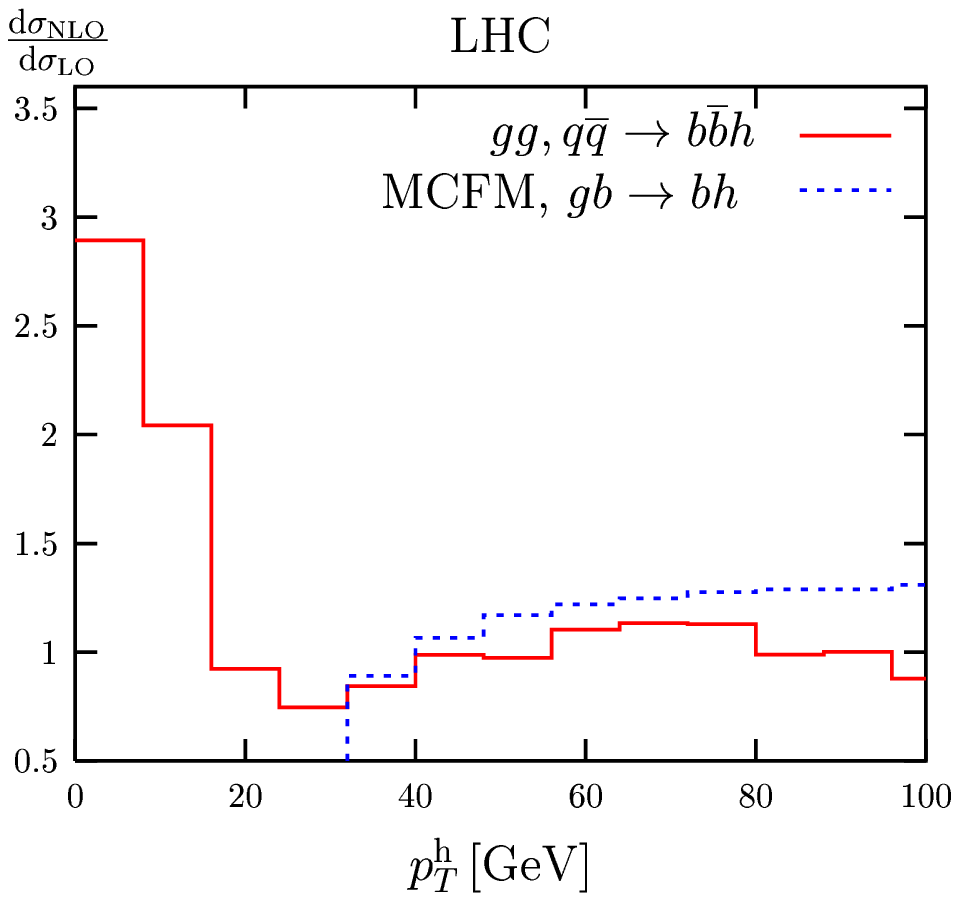}\\
\includegraphics[bb=140 455 430 710,scale=0.42]{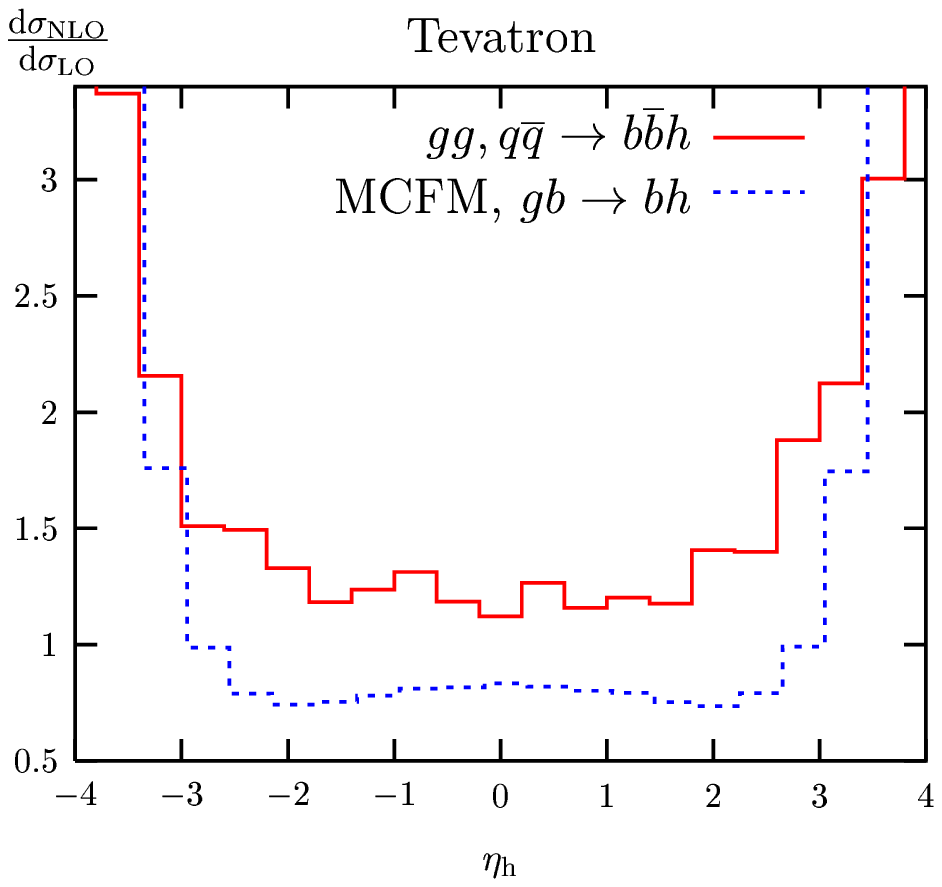}&
\includegraphics[bb=140 455 430 710,scale=0.42]{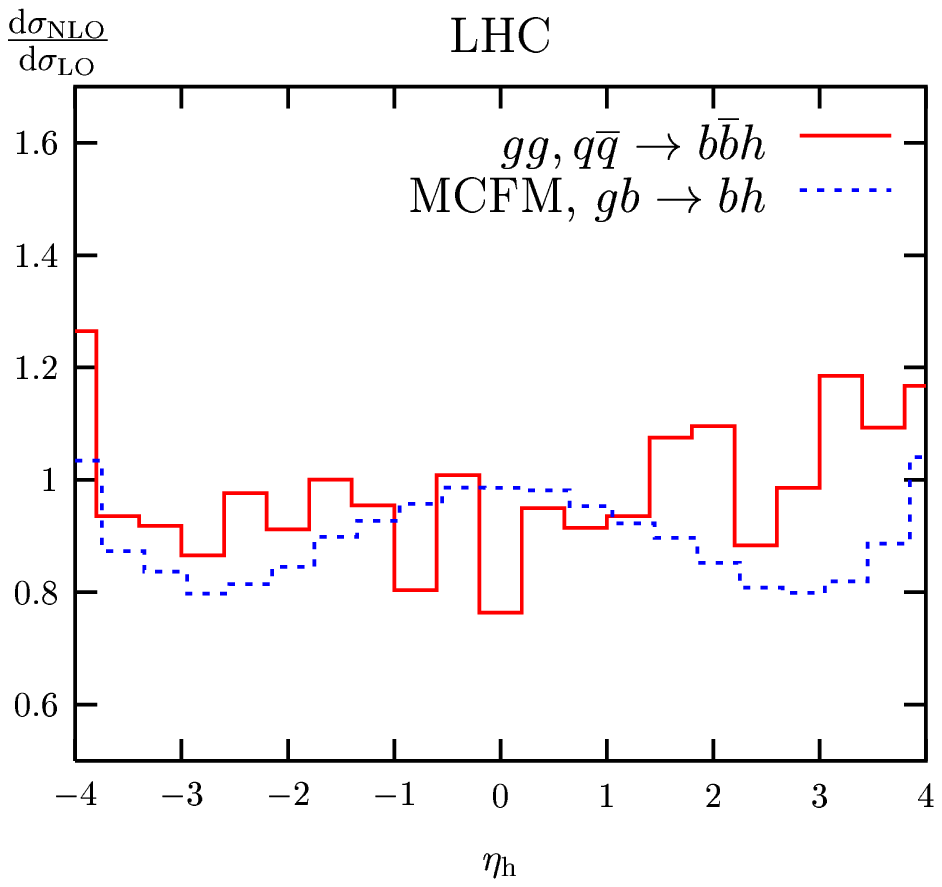}
\end{tabular}
\vspace*{-0.4cm}
\caption[]{
  The ratios of the NLO and LO $p_T^h$ and $\eta_h$ distributions at
  the Tevatron and the LHC for $M_h\!=\!120$~GeV and
  $\mu_r\!=\!\mu_f\!=\! \mu_0/2$. We show the ratios in the 4FNS
  (solid) and 5FNS (dashed).}
\label{fg:dsnlolo}
\end{center}
\end{figure}

Finally, in Figs.~\ref{fg:dsdpt}-\ref{fg:dsnlolo} we compare the
results for the $p_T$ and $\eta$ distributions of the Higgs boson in
both the 4FNS and 5FNS, at the Tevatron and the LHC. We see, in
general, a good agreement between the two schemes, except in regions
of kinematical boundaries. This is particularly dramatic in the
$p_T^h$ distributions where, around $p_T^h\!\simeq\!20$~GeV, a
kinematical threshold causes the 5FNS NLO calculation to be highly
unstable. This instability can be reabsorbed by using a larger bin
size (see inlays), and could therefore be interpreted as a sort of
theoretical \emph{resolution} for the 5FNS.  The instabilities could
be removed by a systematic resummation of threshold corrections
\cite{DelDuca:2003uz,Catani:1997xc}, but this is not implemented in
MCFM. Fig.~\ref{fg:dsnlolo} illustrates the impact of NLO QCD
corrections on $p_T^h$ and $\eta_h$ distributions in terms of a
differential K-factor ($d\sigma_{NLO}/d\sigma_{LO}$). It is
interesting to note that the 4FNS and 5FNS agree at large $p_T^h$ but
they differ substantially at low $p_T^h$. As can be seen in
Fig.~\ref{fg:dsnlolo}, there are regions of $p_T^h$ and $\eta_h$ where
the NLO QCD corrections can considerably affect the shape of the
distributions.

In this letter we have shown how the theoretical predictions for both
total and differential cross sections for $pp,p\bar p\to b(\bar{b})h$
production within a 4FNS and 5FNS are fully compatible within the
existing theoretical errors due to the residual normalization and
factorization scale dependence at NLO in QCD. This is crucial to
experimental searches based on $b\bar{b}h$ production when only one
$b$ quark jet is identified.
\vspace*{-0.8truecm}
\section*{Acknowledgments}
We thank S.~Dittmaier, M.~Kr\"amer, and M.~Spira for comparing results
and J.~Campbell, F.~Maltoni, and S.~Willenbrock for discussions.  We
are particularly grateful to J.~Campbell for his help with the MCFM
program. The work of S.D. (L.R., C.B.J.) and D.W. is supported in part
by the U.S.  Department of Energy under grant DE-AC02-76CH00016
(DE-FG02-97ER41022) and by the National Science Foundation under grant
NSF-PHY-0244875, respectively.

\bibliography{bbh_prl}

\begin{thebibliography}{22}
\expandafter\ifx\csname natexlab\endcsname\relax\def\natexlab#1{#1}\fi
\expandafter\ifx\csname bibnamefont\endcsname\relax
  \def\bibnamefont#1{#1}\fi
\expandafter\ifx\csname bibfnamefont\endcsname\relax
  \def\bibfnamefont#1{#1}\fi
\expandafter\ifx\csname citenamefont\endcsname\relax
  \def\citenamefont#1{#1}\fi
\expandafter\ifx\csname url\endcsname\relax
  \def\url#1{\texttt{#1}}\fi
\expandafter\ifx\csname urlprefix\endcsname\relax\def\urlprefix{URL }\fi
\providecommand{\bibinfo}[2]{#2}
\providecommand{\eprint}[2][]{\url{#2}}

\bibitem[{\citenamefont{Barnett et~al.}(1988)\citenamefont{Barnett, Haber, and
  Soper}}]{Barnett:1987jw}
\bibinfo{author}{\bibfnamefont{R.~M.} \bibnamefont{Barnett}},
  \bibinfo{author}{\bibfnamefont{H.~E.} \bibnamefont{Haber}}, \bibnamefont{and}
  \bibinfo{author}{\bibfnamefont{D.~E.} \bibnamefont{Soper}},
  \bibinfo{journal}{Nucl. Phys.} \textbf{\bibinfo{volume}{B306}},
  \bibinfo{pages}{697} (\bibinfo{year}{1988}).

\bibitem[{\citenamefont{Olness and Tung}(1988)}]{Olness:1987ep}
\bibinfo{author}{\bibfnamefont{F.~I.} \bibnamefont{Olness}} \bibnamefont{and}
  \bibinfo{author}{\bibfnamefont{W.-K.} \bibnamefont{Tung}},
  \bibinfo{journal}{Nucl. Phys.} \textbf{\bibinfo{volume}{B308}},
  \bibinfo{pages}{813} (\bibinfo{year}{1988}).

\bibitem[{\citenamefont{Dicus and Willenbrock}(1989)}]{Dicus:1989cx}
\bibinfo{author}{\bibfnamefont{D.~A.} \bibnamefont{Dicus}} \bibnamefont{and}
  \bibinfo{author}{\bibfnamefont{S.}~\bibnamefont{Willenbrock}},
  \bibinfo{journal}{Phys. Rev.} \textbf{\bibinfo{volume}{D39}},
  \bibinfo{pages}{751} (\bibinfo{year}{1989}).

\bibitem[{\citenamefont{Dicus et~al.}(1999)\citenamefont{Dicus, Stelzer,
  Sullivan, and Willenbrock}}]{Dicus:1998hs}
\bibinfo{author}{\bibfnamefont{D.}~\bibnamefont{Dicus}},
  \bibinfo{author}{\bibfnamefont{T.}~\bibnamefont{Stelzer}},
  \bibinfo{author}{\bibfnamefont{Z.}~\bibnamefont{Sullivan}}, \bibnamefont{and}
  \bibinfo{author}{\bibfnamefont{S.}~\bibnamefont{Willenbrock}},
  \bibinfo{journal}{Phys. Rev.} \textbf{\bibinfo{volume}{D59}},
  \bibinfo{pages}{094016} (\bibinfo{year}{1999}), \eprint{hep-ph/9811492}.

\bibitem[{\citenamefont{Balazs et~al.}(1999)\citenamefont{Balazs, He, and
  Yuan}}]{Balazs:1998sb}
\bibinfo{author}{\bibfnamefont{C.}~\bibnamefont{Balazs}},
  \bibinfo{author}{\bibfnamefont{H.-J.} \bibnamefont{He}}, \bibnamefont{and}
  \bibinfo{author}{\bibfnamefont{C.~P.} \bibnamefont{Yuan}},
  \bibinfo{journal}{Phys. Rev.} \textbf{\bibinfo{volume}{D60}},
  \bibinfo{pages}{114001} (\bibinfo{year}{1999}), \eprint{hep-ph/9812263}.

\bibitem[{\citenamefont{Affolder et~al.}(2001)}]{Affolder:2000rg}
\bibinfo{author}{\bibfnamefont{T.}~\bibnamefont{Affolder}} \bibnamefont{et~al.}
  (\bibinfo{collaboration}{CDF}), \bibinfo{journal}{Phys. Rev. Lett.}
  \textbf{\bibinfo{volume}{86}}, \bibinfo{pages}{4472} (\bibinfo{year}{2001}),
  \eprint{hep-ex/0010052}.

\bibitem[{bbh()}]{bbh_d0}
\bibinfo{note}{D0 collaboration, Results presented at Moriond 2004.}

\bibitem[{\citenamefont{Dittmaier et~al.}(2003)\citenamefont{Dittmaier,
  Kr{\"a}mer, and Spira}}]{Dittmaier:2003ej}
\bibinfo{author}{\bibfnamefont{S.}~\bibnamefont{Dittmaier}},
  \bibinfo{author}{\bibfnamefont{M.}~\bibnamefont{Kr{\"a}mer}},
  \bibnamefont{and} \bibinfo{author}{\bibfnamefont{M.}~\bibnamefont{Spira}}
  (\bibinfo{year}{2003}), \eprint{hep-ph/0309204}.

\bibitem[{\citenamefont{Dawson et~al.}(2004)\citenamefont{Dawson, Jackson,
  Reina, and Wackeroth}}]{Dawson:2003kb}
\bibinfo{author}{\bibfnamefont{S.}~\bibnamefont{Dawson}},
  \bibinfo{author}{\bibfnamefont{C.~B.} \bibnamefont{Jackson}},
  \bibinfo{author}{\bibfnamefont{L.}~\bibnamefont{Reina}}, \bibnamefont{and}
  \bibinfo{author}{\bibfnamefont{D.}~\bibnamefont{Wackeroth}},
  \bibinfo{journal}{Phys. Rev.} \textbf{\bibinfo{volume}{D69}},
  \bibinfo{pages}{074027} (\bibinfo{year}{2004}), \eprint{hep-ph/0311067}.

\bibitem[{\citenamefont{Campbell et~al.}(2003)\citenamefont{Campbell, Ellis,
  Maltoni, and Willenbrock}}]{Campbell:2002zm}
\bibinfo{author}{\bibfnamefont{J.}~\bibnamefont{Campbell}},
  \bibinfo{author}{\bibfnamefont{R.~K.} \bibnamefont{Ellis}},
  \bibinfo{author}{\bibfnamefont{F.}~\bibnamefont{Maltoni}}, \bibnamefont{and}
  \bibinfo{author}{\bibfnamefont{S.}~\bibnamefont{Willenbrock}},
  \bibinfo{journal}{Phys. Rev.} \textbf{\bibinfo{volume}{D67}},
  \bibinfo{pages}{095002} (\bibinfo{year}{2003}), \eprint{hep-ph/0204093}.

\bibitem[{\citenamefont{Campbell et~al.}(2004)}]{Campbell:2004pu}
\bibinfo{author}{\bibfnamefont{J.}~\bibnamefont{Campbell}} \bibnamefont{et~al.}
  (\bibinfo{year}{2004}), \eprint{hep-ph/0405302}.

\bibitem[{\citenamefont{Beenakker et~al.}(2001)\citenamefont{Beenakker,
  Dittmaier, Kr{\"a}mer, Pl{\"u}mper, Spira, and Zerwas}}]{Beenakker:2001rj}
\bibinfo{author}{\bibfnamefont{W.}~\bibnamefont{Beenakker}},
  \bibinfo{author}{\bibfnamefont{S.}~\bibnamefont{Dittmaier}},
  \bibinfo{author}{\bibfnamefont{M.}~\bibnamefont{Kr{\"a}mer}},
  \bibinfo{author}{\bibfnamefont{B.}~\bibnamefont{Pl{\"u}mper}},
  \bibinfo{author}{\bibfnamefont{M.}~\bibnamefont{Spira}}, \bibnamefont{and}
  \bibinfo{author}{\bibfnamefont{P.}~\bibnamefont{Zerwas}},
  \bibinfo{journal}{Phys. Rev. Lett.} \textbf{\bibinfo{volume}{87}},
  \bibinfo{pages}{201805} (\bibinfo{year}{2001}), \eprint{hep-ph/0107081}.

\bibitem[{\citenamefont{Beenakker et~al.}(2003)\citenamefont{Beenakker,
  Dittmaier, Kr{\"a}mer, Pl{\"u}mper, Spira, and Zerwas}}]{Beenakker:2002nc}
\bibinfo{author}{\bibfnamefont{W.}~\bibnamefont{Beenakker}},
  \bibinfo{author}{\bibfnamefont{S.}~\bibnamefont{Dittmaier}},
  \bibinfo{author}{\bibfnamefont{M.}~\bibnamefont{Kr{\"a}mer}},
  \bibinfo{author}{\bibfnamefont{B.}~\bibnamefont{Pl{\"u}mper}},
  \bibinfo{author}{\bibfnamefont{M.}~\bibnamefont{Spira}}, \bibnamefont{and}
  \bibinfo{author}{\bibfnamefont{P.}~\bibnamefont{Zerwas}},
  \bibinfo{journal}{Nucl. Phys.} \textbf{\bibinfo{volume}{B653}},
  \bibinfo{pages}{151} (\bibinfo{year}{2003}), \eprint{hep-ph/0211352}.

\bibitem[{\citenamefont{Reina and Dawson}(2001)}]{Reina:2001sf}
\bibinfo{author}{\bibfnamefont{L.}~\bibnamefont{Reina}} \bibnamefont{and}
  \bibinfo{author}{\bibfnamefont{S.}~\bibnamefont{Dawson}},
  \bibinfo{journal}{Phys. Rev. Lett.} \textbf{\bibinfo{volume}{87}},
  \bibinfo{pages}{201804} (\bibinfo{year}{2001}),
  \eprint[http://arXiv.org/abs]{hep-ph/0107101}.

\bibitem[{\citenamefont{Reina et~al.}(2002)\citenamefont{Reina, Dawson, and
  Wackeroth}}]{Reina:2001bc}
\bibinfo{author}{\bibfnamefont{L.}~\bibnamefont{Reina}},
  \bibinfo{author}{\bibfnamefont{S.}~\bibnamefont{Dawson}}, \bibnamefont{and}
  \bibinfo{author}{\bibfnamefont{D.}~\bibnamefont{Wackeroth}},
  \bibinfo{journal}{Phys. Rev.} \textbf{\bibinfo{volume}{D65}},
  \bibinfo{pages}{053017} (\bibinfo{year}{2002}),
  \eprint[http://arXiv.org/abs]{hep-ph/0109066}.

\bibitem[{\citenamefont{Dawson et~al.}(2003{\natexlab{a}})\citenamefont{Dawson,
  Orr, Reina, and Wackeroth}}]{Dawson:2002tg}
\bibinfo{author}{\bibfnamefont{S.}~\bibnamefont{Dawson}},
  \bibinfo{author}{\bibfnamefont{L.~H.} \bibnamefont{Orr}},
  \bibinfo{author}{\bibfnamefont{L.}~\bibnamefont{Reina}}, \bibnamefont{and}
  \bibinfo{author}{\bibfnamefont{D.}~\bibnamefont{Wackeroth}},
  \bibinfo{journal}{Phys. Rev.} \textbf{\bibinfo{volume}{D67}},
  \bibinfo{pages}{071503} (\bibinfo{year}{2003}{\natexlab{a}}),
  \eprint{hep-ph/0211438}.

\bibitem[{\citenamefont{Dawson et~al.}(2003{\natexlab{b}})\citenamefont{Dawson,
  Jackson, Orr, Reina, and Wackeroth}}]{Dawson:2003zu}
\bibinfo{author}{\bibfnamefont{S.}~\bibnamefont{Dawson}},
  \bibinfo{author}{\bibfnamefont{C.}~\bibnamefont{Jackson}},
  \bibinfo{author}{\bibfnamefont{L.~H.} \bibnamefont{Orr}},
  \bibinfo{author}{\bibfnamefont{L.}~\bibnamefont{Reina}}, \bibnamefont{and}
  \bibinfo{author}{\bibfnamefont{D.}~\bibnamefont{Wackeroth}},
  \bibinfo{journal}{Phys. Rev.} \textbf{\bibinfo{volume}{D68}},
  \bibinfo{pages}{034022} (\bibinfo{year}{2003}{\natexlab{b}}),
  \eprint{hep-ph/0305087}.

\bibitem[{mcf()}]{mcfm}
\bibinfo{note}{J. Campbell and R.K. Ellis, webpage: mcfm.fnal.gov}.

\bibitem[{\citenamefont{Pumplin et~al.}(2002)}]{Pumplin:2002vw}
\bibinfo{author}{\bibfnamefont{J.}~\bibnamefont{Pumplin}} \bibnamefont{et~al.},
  \bibinfo{journal}{JHEP} \textbf{\bibinfo{volume}{07}}, \bibinfo{pages}{012}
  (\bibinfo{year}{2002}), \eprint{hep-ph/0201195}.

\bibitem[{\citenamefont{Stump et~al.}(2003)}]{Stump:2003yu}
\bibinfo{author}{\bibfnamefont{D.}~\bibnamefont{Stump}} \bibnamefont{et~al.},
  \bibinfo{journal}{JHEP} \textbf{\bibinfo{volume}{10}}, \bibinfo{pages}{046}
  (\bibinfo{year}{2003}), \eprint{hep-ph/0303013}.

\bibitem[{\citenamefont{Del~Duca et~al.}(2003)\citenamefont{Del~Duca, Maltoni,
  Nagy, and Trocsanyi}}]{DelDuca:2003uz}
\bibinfo{author}{\bibfnamefont{V.}~\bibnamefont{Del~Duca}},
  \bibinfo{author}{\bibfnamefont{F.}~\bibnamefont{Maltoni}},
  \bibinfo{author}{\bibfnamefont{Z.}~\bibnamefont{Nagy}}, \bibnamefont{and}
  \bibinfo{author}{\bibfnamefont{Z.}~\bibnamefont{Trocsanyi}},
  \bibinfo{journal}{JHEP} \textbf{\bibinfo{volume}{04}}, \bibinfo{pages}{059}
  (\bibinfo{year}{2003}), \eprint{hep-ph/0303012}.

\bibitem[{\citenamefont{Catani and Webber}(1997)}]{Catani:1997xc}
\bibinfo{author}{\bibfnamefont{S.}~\bibnamefont{Catani}} \bibnamefont{and}
  \bibinfo{author}{\bibfnamefont{B.~R.} \bibnamefont{Webber}},
  \bibinfo{journal}{JHEP} \textbf{\bibinfo{volume}{10}}, \bibinfo{pages}{005}
  (\bibinfo{year}{1997}), \eprint{hep-ph/9710333}.

\end{thebibliography}
\end{document}